\documentstyle[12pt]{article}
 1

\begin{document}
\begin{titlepage}
\begin{flushright}
{IOA.295/93}\\
\end{flushright}
\begin{center}
{\bf  RADIATIVE SYMMETRY BREAKING AND THE TOP }\\
{\bf    QUARK MASS IN LOCAL SUPERSYMMETRY}\\
\vspace*{1cm} {\bf G.K. Leontaris} \\
\vspace*{0.3cm}
{\it Theoretical Physics Division} \\
{\it Ioannina University} \\
{\it GR-45110 Greece} \\
\vspace*{0.3cm}
\vspace*{0.5cm}
{\bf ABSTRACT} \\
\end{center}
\noindent

Exact low energy expressions are derived for the top-squark
and higgs masses, taking into account radiative contributions
due to a heavy top quark. Their masses are expressed as
analytic functions of $m_{1/2}$, $m_{3/2}$, $m_t$
and $sin\beta$. Constraints from the radiative symmetry breaking
mechanism are used then, to put lower bounds on the top-quark
mass  $m_t$. In particular, when $m_{3/2}\gg m_{1/2}$, we
obtain the bound $m_t \ge 150sin\beta GeV$, while in the case of
$m_{3/2}\equiv 0$, we obtain $m_t \ge 80sin\beta GeV$.

\vspace*{5cm}
\noindent
\begin{flushleft}
IOA-295/93 \\
July 1993
\end{flushleft}
\vfill\eject
\vfill\eject
\setcounter{page}{1}

It has by now been established that the idea of Grand Unification with the
minimal standard model content, is realised only when supersymmetry is
present down to the $TeV$ scale. On the other hand, local supersymmetry is
necessary in order to solve the gauge hierarchy problem\cite{NLN}.
Therefore, spontaneously broken Supergravity coupled to Grand Unified
Theories, is a promising scenario for the physics beyond the standard
model.

Another nice feature of Supersymmetric theories,
 is the radiative breaking
scenario of the electroweak symmetry\cite{IR,RAD} through the
generalization of the Coleman-Weinberg mechanism\cite{CW}.
The key role in this mechanism is played by the soft SUSY breaking terms
and in patricular, the scalar masses which, as going from the
Unification scale ($M_U$) down to low energies,
get modified  due to large radiative corrections\cite{RAD,EKN}.
The $SU(2)\times U(1)$ symmetry breaking
is obtained if some of the higgs doublets develops a negative
$(mass)^2$. The only radiative corrections which may drive the higgs
$(mass)^2$ negative, arise from the contributions of Yukawa couplings through
the renormalization group running of the scalar masses. Obviously, the
most important contribution of this kind arises from the top-quark Yukawa
coupling. Such contributions have already been estimated in the
literature\cite{RAD,EKN,EKN1,GL}.

Because of their importance in the radiative breaking mechanism,
and in view of the new theoretical predictions and experimental constraints
for the top-mass range, we would like in this letter to present a new
approach and calculte more accurately their effects. We will make a
first attempt to derive analytic expressions for these negative
contributions,  and discuss  the bounds imposed on the top-Yukawa
coupling.

In order to set our formalism and the general framework, we would like to
start from the well known results on the evolution of the top-Yukawa
coupling $\lambda_{top}$, due to renormalization group running from the
Unification point down to low energies.
We will assume  that $\lambda_{top}$ is much bigger
than all other fermion Yukawa couplings. In this case
we may ignore the contributions of
the latter, thus the top-Yukawa coupling differential equation
  may be cast in the form
 \begin{eqnarray}
16\pi^2 \frac{d}{dt}
\lambda_{top}&=&\lambda_{top}(6{\lambda}_{top}
^2-G_U(t))\label{eq:topeq} \end{eqnarray}
The relevant
gauge contribution $G_U$ is given by
\begin{eqnarray}
G_U&=&\sum_{i=1}^3 c_U^i g_i^2(t),\\
g_i^2(t)&=&\frac{g_i^2(t_0)}{1- \frac{b_i}{8\pi^2} g_i^2(t_0)(t-t_0)}
\end{eqnarray}
The $g_i$ are the three gauge coupling constants
of the Standard Model and $b_i$
are the corresponding supersymmetric beta functions. The coefficients
$c_U^i$ are given by
\begin{eqnarray}
\{c_U^i \}_{i=1,2,3} = \left\{ \frac{13}{15},3,\frac{16}{3} \right\}
\end{eqnarray}
The solution of Eq(\ref{eq:topeq}) is
\begin{eqnarray}
\lambda_{top}(t)&=& \lambda_{top}(t_0)\xi(t)^6\gamma_U(t)\
\label{eq:ltop}
\end{eqnarray}
where\cite{DL,GG}
\begin{eqnarray}
\gamma_U(t)&=& \exp({-\frac{1}{16\pi^2}\int_{t_0}^{t} G_U(t) \,dt})
\nonumber\\
&=& \prod_{j=1}^3 \left(1- \frac{b_{j,0}
\alpha_{j,0}(t-t_0)}{2\pi}
\right)^{c_U^j/2b_j},
\end{eqnarray}
and
\begin{eqnarray}
\xi(t)&=&\exp({\frac{1}{16\pi^2}\int_{t_0}^{t} \lambda_{top}^2(t) \,dt})
\nonumber\\
&=& \left( 1-\frac{6}{8\pi^2}\lambda_{top}^2(t_0)
\int_{t_0}^{t} \gamma_U^2(t)\,dt \right)^{-1/12}\label{eq:ksi}.
\end{eqnarray}

Let us now turn our attention
in the scalar masses. As we have already mentioned, we
are interested in those low energy scalar masses which are affected by
the top - Yukawa contributions. These are the top-squarks $m_{\tilde t_L},
m_{\tilde t_R}$, the  higgs mass $m_{\bar h}$, (where $\bar h$ is
 the higgs which gives masses to the up-quarks) and
the trilinear scalar coupling parameter $A$.
Let us denote $m_{\tilde t_L}\equiv \tilde m_1$,
$m_{\tilde t_R}\equiv \tilde m_2$, and
$m_ {\bar h}\equiv \tilde m_3$.
Then, at any scale
$t=ln{\mu}$, we can write the following general formula for the squark
 and $\bar h$-higgs masses which couple to the top-Yukawa coupling
 \begin{eqnarray}
\tilde m^2_n(t)=\alpha_nm_{3/2}^2+C_n(t)m_{1/2}^2
-n\Delta^2_A(t)-n\Delta^2_{\tilde m}(t)
\label{eq:YCn}
\end{eqnarray}
In the above,
\begin{eqnarray}
\Delta^2_A(t)+\Delta^2_{\tilde m}(t)=\int_{t}^{t_0}
\frac{\lambda_{top}^2(t^{\prime})}{8\pi^2}
\left(A^2(t)+\sum_{j=1}^3\tilde m^2_j(t)\right)\,dt^{\prime}
\label{eq: Delta}
\end{eqnarray}
where with $\Delta_A(t)$ we have denoted the radiative corrections due to
the trilinear scalar coupling which in general depends on the scale
parameter $t=ln\mu$\cite{RAD}.
 $C_n(t)=C_Q,C_U,C_L$, are calculable coefficients \cite{CK}
 which represent gauge corrections, while
$\alpha_n\equiv \alpha_Q=\alpha_U=\alpha_L=1$ in the
minimal case where the scalars are in a flat manifold\cite{CFGP}.
Thus, Eq(\ref{eq:YCn}), can be transformed to an integral equation of
Volterra type which can be solved exactly. We start by summing up all
the scalar masses containing the Yukawa corrections. Thus we get
\begin{eqnarray}
\sum_{n=1}^3\tilde m^2_n(t)=
\sum_{n=1}^3\alpha_nm_{3/2}^2+
\sum_{n=1}^3C_n(t)m_{1/2}^2-\sum_{n=1}^3n\Delta_A^2(t)
-\sum_{n=1}^3n\Delta^2_{\tilde m}(t)
\label{eq: sumn}
\end{eqnarray}
Let us define
\begin{eqnarray}
u(t)=\sum_{n=1}^3\tilde m^2_n(t)\nonumber\\
u_0(t)=\sum_{n=1}^3\alpha_nm_{3/2}^2+
\sum_{n=1}^3C_nm_{1/2}^2-\sum_{n=1}^3n\Delta_A^2(t)\label{eq: def}\\
C=-\frac{1}{8\pi^2}\sum_{n=1}^3n\nonumber
\end{eqnarray}
 Using the
definitions in Eqs(\ref{eq: def}) above,
 Eq(\ref{eq: sumn}) can be cast in a standard form
\begin{eqnarray}
u(t)=
u_0(t)+C\int_{t}^{t_0}
{dt^{\prime}\lambda_{top}^2(t^{\prime})u(t^{\prime})}\label{eq: ut}
\end{eqnarray}
 The above equation can be solved easily.
Define \begin{eqnarray}
f(t)=C\int_{t}^{t_0}
{dt^{\prime}\lambda_{top}^2(t^{\prime})u(t^{\prime})}\label{eq: ft}
\end{eqnarray}
Differentiating Eq(\ref{eq: ft}), and substituting back Eq(\ref{eq:
ut}), we obtain
\begin{eqnarray}
\frac{\,d f(t)}{ u_0(t)+f(t)}=-C\lambda_{top}^2(t)
\end{eqnarray}
The solution of the above is
\begin{eqnarray}
 f(t)&=&-q(t)^{-1}C\int_{t_0}^{t}q(t^{\prime})u_0(t^{\prime})
\lambda_{top}^2(t^{\prime})\,dt^{\prime}
\label{eq:sft}
\end{eqnarray}
where the {\it integrating factor } $q=q(t)$, is given by
\begin{eqnarray}
q&=&exp\{C\int_{t_0}^{t}\lambda_{top}^2(t^{\prime})\,dt^{\prime}\}
\nonumber\\
&=&exp\{-12ln{\xi(t)}\}\label{eq:qt}\\
&=&\xi^{-12}
\end{eqnarray}
where $\xi$ is given again by the integral (\ref{eq:ksi}).
Now, the solution for $u(t)$ can be given by substituting Eq(\ref{eq:sft})
into (\ref{eq: ut}). The corrections $\Delta^2(t)$, can also be expressed in
terms of the function $f(t)$,
\begin{eqnarray}
\Delta^2_{\tilde m}(t)&=&-\frac{1}{6}f(t)
 \label{eq:sdt}
\end{eqnarray}
Substituting the above results in Eq(\ref{eq:YCn}), we find that
 the scalar masses are given by
\begin{eqnarray}
\tilde m ^2 _n(t)&=&\alpha_nm_{3/2}^2+C_n(t)m_{1/2}^2
{}-{}n\delta_A^2(t){}-{}n\delta_m^2(t)
\label{eq: smn}
\end{eqnarray}
where
\begin{eqnarray}
\delta_m^2(t)&=&\xi(t)^{12}
\int_{t_0}^{t}\xi(t^{\prime})^{-12}
\left(C\lambda_{top}^2(t^{\prime})dt^{\prime}\right)
\nonumber\\
{}&\times &\frac{1}{6}(\sum_{n=1}^3\alpha_nm_{3/2}^2+
\sum_{n=1}^3C_n(t^{\prime})m_{1/2}^2),\label{eq:dm}
\end{eqnarray}
and,
\begin{eqnarray}
\delta_A^2(t)&=&\Delta_A^2(t)\nonumber\\
{}&-&\xi(t)^{12}
\int_{t_0}^{t}\xi(t^{\prime})^{-12}
\left(C\lambda_{top}^2(t^{\prime})dt^{\prime}\right)
\Delta_A^2(t^{\prime}).\label{eq: da}
\end{eqnarray}

The above expressions can be simpified by substituting
$\lambda_{top}^2$ from Eq(\ref{eq:ltop}) into the
integrals . Taking also into
account that the top-mass is given by the formula
$m_t=\lambda_{top}\frac{\upsilon}{\sqrt 2}sin\beta$
where $\upsilon =246GeV$, we may write $\delta_m^2(t)$, as follows
\begin{eqnarray}
\delta_m^2(t)=\left(\frac{m_t}{2\pi \upsilon\gamma_U sin\beta}\right)^2
\times (3m_{3/2}^2I+m_{1/2}^2J)\label{eq:dm1}
\end{eqnarray}
where $I,J$, are integrals containing only functions of gauge couplings,
 i.e.
\begin{eqnarray}
I&=&\int_{t}^{t_0}\,dt^{\prime}  \gamma_U^2(t^{\prime})\\
J&=&\int_{t}^{t_0}\,dt^{\prime}  \gamma_U^2(t
^{\prime})C(t^{\prime} )\label{eq: IJ}
\end{eqnarray}
with $C(t)\equiv \sum_{n=1}^3C_n(t^{\prime})$\cite{RAD}.
 A similar form may be obtained for $\delta_A^2(t)$.

As an application, let us calculate the  $(mass)^2$-parameter  of the
 $\bar h$-higgs.
We start with the integral (\ref{eq:dm}).
The integrals $I,J$ can be easily performed numerically. Thus
for $t\sim lnm_t$, with   $m_t$ in the experimentally allowed region,
 we get $I\approx 112$, and $J\approx 680$.

The trilinear coupling $A$ is given by\cite{RAD}
$A(t)=(A_0m_{3/2}+C_A(t)m_{1/2})/(1-12ln\xi)$.
For our present perposes let us assume an average value for $C_A(t)$
and ignore the $t$-dependence.
Now the integral can be performed rather
easily by observing that the various $t$-dependent quantities  can finally be
expressed as functions of $\xi$. Indeed, first we observe that
($\xi(t^{\prime})\equiv \xi^{\prime}$),
\begin{eqnarray}
C\lambda_{top}^2(t^{\prime})dt^{\prime}=d(-12\xi^{\prime})
\end{eqnarray}
Under the previous assumption, the integral $\Delta_A^2$ is
\begin{eqnarray}
\Delta_A^2(t)&=&\frac{1}{6}Q_0^2\int_{0}^{-12ln\xi}
\frac{{d(-12\xi^{\prime})}}{(1-12\xi^{\prime})^2}
\end{eqnarray}
with $Q_0=(A_0m_{3/2}+C_Am_{1/2})$. Thus, the quantity $\delta_A^2(t)$
becomes
\begin{eqnarray}
\delta_A^2(t)=\frac{1}{6}Q_0^2\left(\int_{0}^{-12ln\xi}
\frac{1}{(1+x)^2}\,dx-
\xi(t)^{12}\int_{0}^{-12ln\xi}\frac{xe^x}{1+x}\,dx\right)
\end{eqnarray}
{}For $\mu\approx  O(m_t)$ as previously, a correct bottom mass prediction
requires \cite{GG,DLT} $\xi\approx 0.81$, thus $\delta_A^2\approx
2.52\times 10^{-2}Q_0^2$, which is negligible compared to $\delta_m^2$
contribution.

Substituting (\ref{eq:dm1}) back to the formula (\ref{eq: smn}),
in the case $m_{3/2}\equiv 0$ for example, and requiring $m_ {\bar h}^2<0$,
we obtain the top-mass limit  $m_t\ge 80 sin\beta GeV$, while in the case
 $m_{1/2}= 0$, we obtain  $m_t\ge 150 sin\beta GeV$. No upper bound in
obtained for $m_t$ from the other scalar masses.
A more detailed analysis will be presented in a future publication.

{\it I would like to thank C. Kounnas for useful discussions.}

\end{document}